\documentclass[%
 reprint,
superscriptaddress,
 amsmath,amssymb,
 aps,
prb,
]{revtex4-2}

\usepackage{gensymb} 
\usepackage{graphicx}
\usepackage{dcolumn}
\usepackage{bm}
\usepackage{ulem}

\usepackage{xcolor} 

\begin{document}


\title{Phase transition between two different orientations of the Q phase in the NaNbO$_3$ thin film}

\author{A. V. Pavlenko}
\affiliation{Institute of Physics, Southern Federal University, 344090 Rostov-on-Don, Russia}
\affiliation{Southern Scientific Center, Russian Academy of Sciences, 344006 Rostov-on-Don, Russia}
\author{D. V. Stryukov}%
\affiliation{Southern Scientific Center, Russian Academy of Sciences, 344006 Rostov-on-Don, Russia}

\author{M. V. Vladimirov}%
\author{A. E. Ganzha}
\author{S. A. Udovenko} 
\affiliation{Peter the Great Saint-Petersburg Polytechnic University, 195251 St. Petersburg, Russia}

\author{Anjana Joseph}%
\affiliation{Jawaharlal Nehru Centre for Advanced Scientific Research, 560064 Bangalore, India}

\author{Janaky Sunil}%
\affiliation{Jawaharlal Nehru Centre for Advanced Scientific Research, 560064 Bangalore, India}

\author{Chandrabhas Narayana}%
\affiliation{Jawaharlal Nehru Centre for Advanced Scientific Research, 560064 Bangalore, India}

\author{R. G. Burkovsky}%
\affiliation{Peter the Great Saint-Petersburg Polytechnic University, 195251 St. Petersburg, Russia}

\author{I. P. Raevski}%
\affiliation{Institute of Physics, Southern Federal University, 344090 Rostov-on-Don, Russia}

\author{N. V. Ter-Oganessian}
 \email{teroganesyan@sfedu.ru}
\affiliation{Institute of Physics, Southern Federal University, 344090 Rostov-on-Don, Russia}%

\date{\today}

\begin{abstract}
Temperature evolution of dielectric response, atomic structure, and lattice dynamics in thin film of sodium niobate in the epitaxial NaNbO$_3$/SrRuO$_3$/(001)MgO heterostructure is studied by dielectric measurements, x-ray diffraction, and Raman spectroscopy. It is found that at room temperature NaNbO$_3$ is in ferroelectric state, whereas the temperature-dependent dielectric constant experiences a broad maximum at 440~K on cooling and at 500~K on heating and reveals a diffuse phase transition. Reciprocal space mapping shows the presence of both anti-phase and in-phase tilting of oxygen octahedra. The temperature dependence of the M-point reflections suggests reorientation of the in-phase octahedra tilting axis from being parallel to the substrate at room temperature to perpendicular orientation at high temperatures.  The temperature evolution of the shape of the Raman spectra reveal the decrease of the number of constituting peaks on heating. These results are interpreted as indicating a temperature-driven transition between two different orientations of the bulk ferroelectric Q phase with respect to the interface, namely between the state with electric polarization pointing at $\approx45\degree$ to the normal at room temperature to the state with polarization parallel to the interface above the transition. Transitions of this kind can be anticipated from theoretical considerations, while the experimental evidences of such are yet scarce.
\end{abstract}

\keywords{NaNbO$_3$, thin films, electric polarization, crystal structure, phase transition, Raman spectroscopy}

\maketitle


\section{Introduction\label{sec:Intro}}

Sodium niobate, NaNbO$_3$ (NNO), has been studied since several decades. Because of the presence of at least six temperature-induced phase transitions at ambient pressure, NaNbO$_3$ is known as arguably the most complex perovskite, which makes this compound methodologically interesting~\cite{Megaw_Seven_Phases_1974,Lines_Glass_BOOK}. Sodium niobate has ferroelectric ground state and room temperature antiferroelectric (AFE) phase, which, combined with its lead-free composition, makes it also promising for energy storage, electrocaloric, electromechanical, and other applications~\cite{Palneedi_EnergyStorage_2018,Wu_Potassium_Sodium_2015,Chauhan_Anti_Ferroelectric_Ceramics_2015,ZUo_NNO_BT_2016,Zho_NNO_Appl_2019,Zhang_NNO_relaxor_Appl_2021}.

Seven phases are commonly assumed in NaNbO$_3$ with more or less well-established crystal symmetry, however subtle differences between some phases still lead to de\-ba\-tes in literature~\cite{Mishra_NaNbO3_2007,Peel_Elusive_R_S_phases_2012,DARLINGTON_1999}. The low-temperature rhombohedral ferroelectric phase N, that is stable below 173~K, has space group $R3c$. Above this temperature centrosymmetric phases are stable: P (sp. gr. $Pbcm$, 173~K -- 633~K), R (sp. gr. $Pmmn$, 633~K -- 753~K), S (sp. gr. $Pmmn$, 753~K -- 793~K), T$_1$ (sp. gr. $Ccmm$, 793~K -- 848~K), T$_2$ (sp. gr. $P4/mbm$, 848~K -- 913~K), and U (sp. gr. $Pm\bar{3}m$, above 913~K). These phase transition temperatures naturally vary slightly from one work to another, except for the N -- P phase transition temperature, which strongly depends on the quality of the crystal, the presence of defects and impurities~\cite{Raevskaya_Quantum_NNO_2008}. Some authors also conclude on the coexistence of phases, e.g., the coexistence of the N and P phases over a wide temperature range between 12 and 280~K~\cite{Mishra_NaNbO3_2007,Jiang_2013}.

In some studies of bulk NaNbO$_3$, in the temperature region of the P phase, ferroelectric Q phase with $P2_1ma$ symmetry is observed instead~\cite{Lefkowitz_1966}. This phase can appear in NaNbO$_3$ samples with defects or slight variations in composition, but is also induced by application of electric field, which confirms the AFE nature of the P phase~\cite{Chen_TEM_PSS_1_1988,Chen_TEM_PSS_2_1988,Shakhovoy_2012}. The Q phase was also reported in NaNbO$_3$ powders with particle sizes below 400~nm and ceramics with fine grains~\cite{Shiratori_SizeEffect_2005,Koruza_SizeEffect_Q_Phase_2010,KORUZA_Grain_Size_2017}. It has to be noted, that reversible P -- Q electric field-induced transition was only observed in high quality single crystals~\cite{Cross_1955,Cross_1958,Ulinzheyev_1990}, which means that in nominally pure NaNbO$_3$ after removal of electric field the Q phase does not revert back do the P phase but is believed to be pinned by defects or due to rather high potential barrier between the two phases. This also evidences that both phases have similar thermodynamic potential values and that of the P phase is slightly more preferable.

Based on the aforementioned facts, it is commonly believed that several structural instabilities resulting in the variety of crystal structures come into play in sodium niobate and the fragile balance between them determines the complex sequence of the observed phase transitions. Therefore, the analysis of the principal distortion modes that result in the observed phases is crucial for the proper understanding of the phase transition sequence. Such structural mode analysis of NaNbO$_3$ was first performed by Cochran and Zia~\cite{Cochran_Zia_1968} and recently extended by Tol{\'{e}}dano and Khalyavin~\cite{Toledano_AF_2019}. The authors conclude that the driving modes belong to several points in the Brillouin zone including $\Gamma$, R, M, and T. These modes, which appear in different combinations in different phases of NaNbO$_3$, have the largest amplitudes and determine their symmetry, whereas other modes are induced in improper way.

Thin films of sodium niobate attract interest of researchers because of possibility of tailoring its properties for technological applications~\cite{Setter_Applications_2006,Sando_Films_Optical_Appl_2018,Burns_KNNO_films_2021}.
Owing to the mechanical constraints imposed by the substrate, size effects, the appearance of dead layers, etc., thin films of NaNbO$_3$ may show physical properties and phase transitions sequences that substantially differ from those of the bulk due to the different influences of these factors on the aforementioned modes.

Thin films of NaNbO$_3$ have been fabricated by different methods including pulsed laser~\cite{Yuzyuk_MgO_film_2010,Yamazoe_NNO_STO_2012,Oda_NaNbO3_MgO_2008,Yamazoe_NaNbO3_SrTiO3_2009,Tyunina_MgO_2009} and metal organic chemical vapor deposition~\cite{Cai_NdGaO3_2014,Duk_TbScO3_2013,Schwarzkopf_Orth_Substr_2014}, magnetron~\cite{Mino_NaNbO3_MgO_2007} and radio-frequency cathode sputtering~\cite{Pavlenko_NaNbO3_2020}. To characterize the structure and properties of the films, x-ray $\theta$/$2\theta$ diffraction and reciprocal space mapping~\cite{Cai_NdGaO3_2014,Duk_TbScO3_2013,Schwarzkopf_Orth_Substr_2014,Mino_NaNbO3_MgO_2007}, transmission electron microscopy~\cite{Yamazoe_NNO_STO_2012,Oda_NaNbO3_MgO_2008}, Raman spectroscopy~\cite{Yuzyuk_MgO_film_2010}, temperature-dependent dielectric studies~\cite{Cai_NdGaO3_2014,Yamazoe_NNO_STO_2012,Tyunina_MgO_2009}, electric polarization loops measurement~\cite{Mino_NaNbO3_MgO_2007,Oda_NaNbO3_MgO_2008,Yamazoe_NaNbO3_SrTiO3_2009}, and piezoelectric force microscopy~\cite{Duk_TbScO3_2013,Schwarzkopf_Orth_Substr_2014,Yamazoe_NNO_STO_2012,Oda_NaNbO3_MgO_2008} methods are commonly employed. With the help of the aforementioned techniques the following results, that are pertinent to the present study, were obtained.

NaNbO$_3$/SrRuO$_3$/SrTiO$_3$ films with differently oriented substrates [(001), (110), and (111)] showed ferroelectric $P$-$E$ polarization loops with the largest polarization in the (110) oriented film~\cite{Yamazoe_NaNbO3_SrTiO3_2009}. A fourfold multiplication of the unit cell perpendicular to the (001)SrTiO$_3$ substrate was found by Yamazoe {\it et al.}~\cite{Yamazoe_NNO_STO_2012}, which allowed the authors to conclude on the $Pbma$ symmetry of the film that was supported by the antiferroelectric-like dielectric anomaly at 645~K. The films with (110) and (111) orientations of SrTiO$_3$ showed a broad dielectric anomaly around 673~K and a twofold multiplication of the unit cell, which let the authors claim the film to have the $P2_1ma$ symmetry (Q phase)~\cite{Yamazoe_NNO_STO_2012}. Sodium niobate films were also deposited on several other perovskite-like (110)-oriented ortho\-rhom\-bic substrates, e.g., NdGaO$_3$, TbScO$_3$, DyScO$_3$, GdScO$_3$ ~\cite{Cai_NdGaO3_2014,Duk_TbScO3_2013,Schwarzkopf_Orth_Substr_2014}. Deposited on such substrates, NaNbO$_3$ exhibits slightly different in-plane lattice parameters $a$ and $b$, which results in in-plane anisotropic dielectric properties, whereas the crystal symmetry was found to be $Pmc2_1$ ($=P2_1ma$, Q phase) in some works~\cite{Cai_NdGaO3_2014,Schwarzkopf_Orth_Substr_2014}.

Several works reported on the synthesis of NaNbO$_3$ films on MgO, which is another frequently used substrate. The type of the substrate can substantially influence the properties of the film despite the same intermediate layer. Thus, NaNbO$_3$/SrRuO$_3$ grown on (001)SrTiO$_3$ or (001)MgO can in principle show different properties. NaNbO$_3$/SrRuO$_3$/(001)MgO films with K-Ta-O or Pt buffers obtained by pulsed laser deposition and RF magnetron sputtering, respectively, were reported to show different pseudotetragonal unit cell parameters~\cite{Oda_NaNbO3_MgO_2008,Mino_NaNbO3_MgO_2007}. The former film showed superstructure along the $c$-axis direction with fourfold multiplication of the lattice constant and both films were found to be ferroelectric with remanent polarizations of 6.4 and 20~$\mu$C/cm$^2$, respectively. Using micro-Raman spectroscopy Yuzyuk {\it et al.} reported on the Q phase stability up to $\sim$600~K in a 250~nm thick NaNbO$_3$/(La$_{0.5}$,Sr$_{0.5}$)CoO$_3$/(001)MgO film~\cite{Yuzyuk_MgO_film_2010}, whereas Tyunina and Levoska also reported a dielectric anomaly around 330~$^\circ$C~\cite{Tyunina_MgO_2009}.

Thin films of NaNbO$_3$ were studied theoretically by Di\'{e}guez {\it et al.}~\cite{Dieguez_Strain_NaNbO3_2005} using density functional theory (DFT) calculations, who determined the direction of electric polarization depending on misfit strain. This approach, which follows the original works of Pertsev {\it et al.}~\cite{Pertsev_Films_1998,Pertsev_films_1999_equil,Pertsev_Films_2001} allows calculating 
phase diagrams in temperature-epitaxial strain coordinates. These diagrams can generally contain phases with the same symmetry that differ only in orientation of the electric polarization. In case of an unstrained bulk crystal these phase states can actually become different domains of the same phase. However, in the case of epitaxial conjugation of the film with the substrate resulting in misfit strain, they can become different phases, resulting in a possibility of phase transition between them with respective anomalies in dielectric constant as, for example, predicted for the phase transition in PbTiO$_3$ between the $c$-phase and the polydomain $a_1/a_2/a_1/a_2$ state~\cite{Pertsev_Films_2001}.

The phase transitions in NNO are due to interplay of several instabilities, which requires taking into account several order parameters that are not considered in the aforementioned approach. Recently, Patel {\it et al.}~\cite{Pattel_2021} also considered the effect of epitaxial strain on (001)NaNbO$_3$ films using DFT method and obtained a phase diagram by comparing the energies of different structures of NNO. The diagram was found to include the monoclinically distorted N phase, the Q phase, and the polar phase stemming from the bulk P structure, in which electric polarization additionally appears, thus, lowering the symmetry from $Pbcm$ to $Pca2_1$. It has to be noted, though, that thorough comparison of theoretical results with experimental findings is still lacking, which is to some extent limited by the frequent absence of reports of the experimentally observed lattice parameters or difficulties in determination of the appearing crystal structure.

Summarizing the main results on the temperature-dependent structural and dielectric properties of NNO films, one can note the appearance of ferroelectric phases at room temperature, observation of two- or fourfold unit cell multiplication in one of the crystal axis directions, occurrence of a maximum in the temperature-dependent dielectric constant. It has to be noted, though, that usually the crystal symmetry of the NNO film, i.e., the Q or the P phase, is established only upon the presence of two- or fourfold multiplication of the pseudocubic unit cell. Since the NNO films with fourfold multiplication in some cases show electric polarization, it remains unclear, for example, whether its structure is related to the aforementioned $Pca2_1$ phase~\cite{Pattel_2021}. Another important open question is: What is the origin of the temperature-induced phase transition manifested in the maximum of the temperature-dependent dielectric constant and how it is related to the dielectric constant maximum at the P-R transition in the bulk? To the best of our knowledge, these questions have not been addressed by any structural temperature-dependent studies, which could shed light on the relation between the crystal structures and phase transition sequence in the film and the bulk forms of NaNbO$_3$.

Motivated by these problems, in this work we study the phase transition in sodium niobate in the NaNbO$_3$/SrRuO$_3$/(001)MgO heterostructure. Using complementary methods including temperature-dependent dielectric, structural, and Raman spectroscopy studies we find a broad phase transition above room temperature. This phase transition is manifested in a broad maximum of the dielectric constant, reorientation of the octahedra tilting axes, and the change in the number of observed Raman modes. The results suggest that this transition is a phase transition between two different orientations of the bulk Q phase with differently oriented polarizations. Similar phase transitions have been actively studied earlier theoretically using phenomenological models~\cite{Pertsev_Films_1998,Pertsev_films_1999_equil,Pertsev_Films_2001,Pertsev_PZT_2003}. In particular, strain-induced transitions were predicted in PbTiO$_3$ and BaTiO$_3$ between states with different orientations of electric polarization, which formally have the same symmetry but should be treated as different phases. The experimental evidence of such temperature-induced phase transitions, which we observe in our work, is still limited.

\section{Experimental}

The films of NaNbO$_3$ and SrRuO$_3$, which was used as the bottom electrode, in the NaNbO$_3$/SrRuO$_3$/(001)MgO [NNO/SRO/(001)MgO] heterostructure were prepared by RF sputtering method using the "Plasma 50 SE" apparatus. Single-crystalline MgO(001) with thickness of 0.5~mm (MTI Corp.) was employed as substrate. The temperature of the substrates in the chamber before switching on the RF discharge was 673~K and 783~K after switching on, the RF power was 120~W, whereas the distance between the target and the furnace with the substrate was 15~mm. First, SrRuO$_3$ was sputtered on MgO to obtain SrRuO$_3$/(001)MgO, which was followed by NaNbO$_3$ sputtering resulting in the sought heterostructure. For the dielectric and Raman spectroscopy measurements the thicknesses of the NNO and SRO films were 750 and 150~nm, respectively, which was suitable for dielectric studies and ensured proper shielding of MgO substrate in Raman measurements, whereas for the x-ray diffraction measurements the thicknesses were 150 and 20~nm, respectively, which ensured the absence of noticable reflections of SRO in reciprocal space mapping.

To conduct dielectric measurements in the direction perpendicular to the film plane, electrodes were deposited on the free surface of the film by thermal evaporation of Al in vacuum through a mask with holes with a diameter of 180-200~$\mu$m, whereas SrRuO$_3$ acted as the lower electrode. The area of electrodes was measured using the Keyence VK-9700 3D microscope (Joint Center for Scientific and Technological Equipment, Scientific Center of the Russian Academy of Sciences). The temperature dependences of the dielectric constant at frequencies of 10$^4$ and 10$^5$~Hz with an amplitude of 0.04~V were obtained using an LCR 4263B meter (Agilent Technologies). The capacitance-voltage $C(U)$ and $P(E)$ dependencies were measured using a TF Analyzer 2000 (Center for Collective Usage of the Institute of Physics, Southern Federal University, Rostov-on-Don, Russia).

X-Ray diffraction (XRD) measurements were carried out on SuperNova (Rigaku) X-Ray diffractometer (Research and Education center “Physics of Nanocomposite Materials”, Peter the Great St. Petersburg Polytechnic University). To characterize sodium niobate film in the NNO/SRO/MgO heterostructure single crystal diffraction method with grazing incidence geometry was used. The incident radiation (Cu K$\alpha$) was scattered on the heterostructure, and the scattered radiation was registered on the detector (Atlas CCD Detector, Agilent Technologies). Two diffraction experiments were performed. In the first experiment $\omega$, $\theta$, and $\kappa$ angles were 72$\degree$, 60$\degree$, and -134$\degree$, respectively, while $\phi$ was changing from 0$\degree$ to 180$\degree$ with a step of 0.25 (1.00)$\degree$. Detector distance was 50~mm and exposure time was 10 (6) seconds. In the second experiment $\omega$, $\theta$, and $\kappa$ angles were 70$\degree$, 70$\degree$, and -105$\degree$, respectively, while $\phi$ was changing from -70$\degree$ to 200$\degree$ with a step of 1$\degree$. Detector distance was 50~mm and exposure time was 10 seconds. The frames obtained from the experiments were used to construct reciprocal space maps.

Raman spectroscopic measurements were conducted using LabRam HR Evolution micro-Raman spectrometer, which has an assembly of 532~nm Nd:YAG solid state four level laser, an 800~mm focal length monochromator and a Peltier-cooled CCD detector (Jawaharlal Nehru Centre for Advanced Scientific Research, Bangalore, India). Each spectrum was collected in 180$\degree$ backscattering geometry using a 50x objective, with an average acquisition time of 1 minute and an incident power of $\sim20$~mW. The LabSpec6 software recorded the spectra in spectral range 100 -- 1000~cm$^{-1}$, with the spectral resolution of around 1~cm$^{-1}$ for the grating of 1800 groves per mm$^{-1}$. The temperature-dependent Raman studies were done from 298~K to 850~K, employing the Linkam THMS 600 cryostage with a temperature controller (Linkam TMS 94), which maintains the set temperature with an accuracy of $\pm0.1$~K.

\section{Results}

\subsection{Dielectric studies}

\begin{figure}[t]
\centering
\includegraphics[width=\columnwidth]{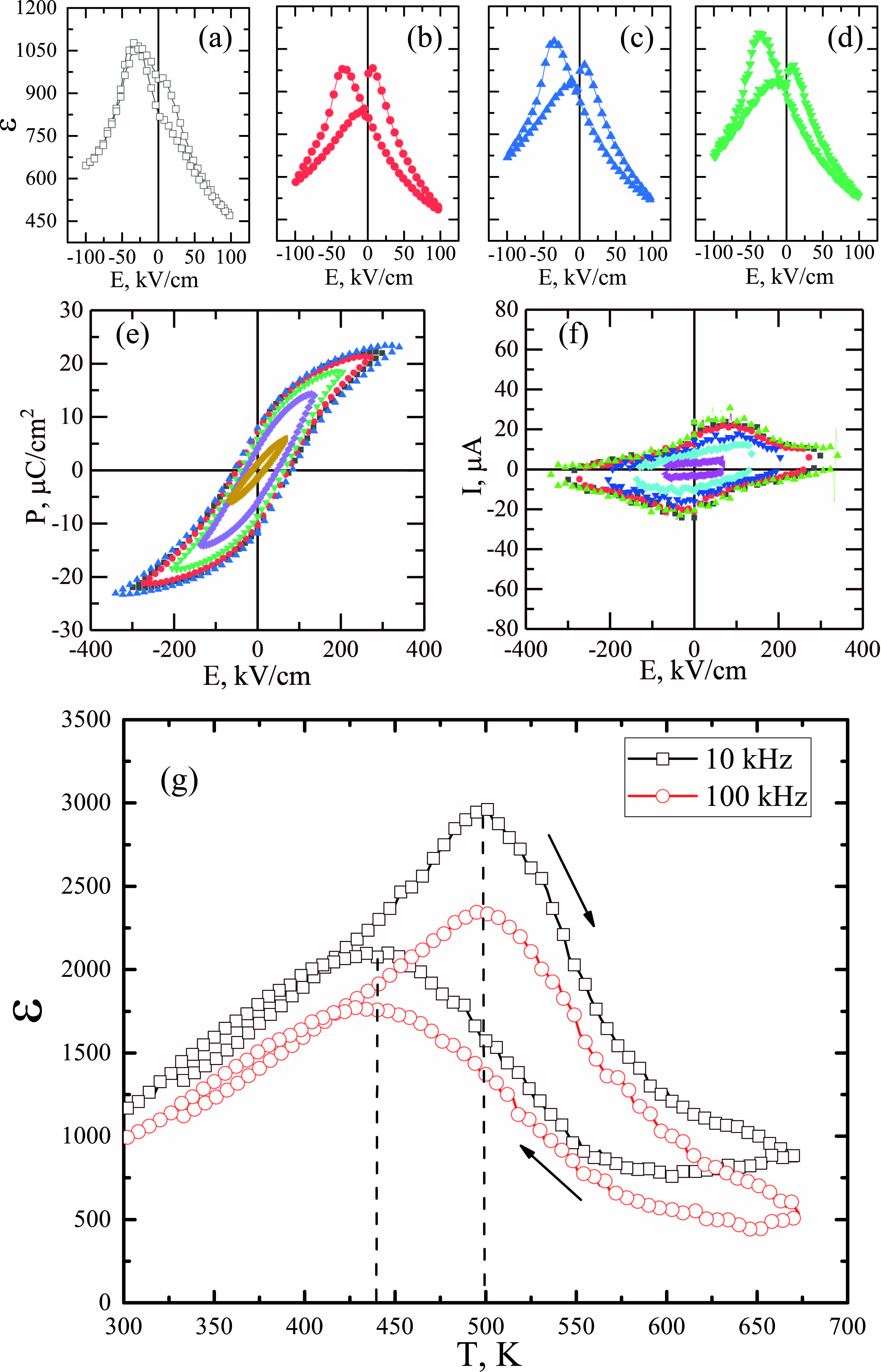}%
\caption{(a) -- (d) Dielectric constant dependencies on electric field $\varepsilon(E)$. The frequency of the measurement electric field with the amplitude 40~mV is 100~kHz and the cycle frequencies are: (a) -- $f_{\rm ext}$=1~Hz, (b) -- $f_{\rm ext}$=5~Hz, (c) -- $f_{\rm ext}$=10~Hz, (d) -- $f_{\rm ext}$=40~Hz. Electric polarization $P(E)$ (e) and electric current $I(E)$ (f) as functions of applied electric field at room temperature and the frequency of 1~kHz. (g) Temperature dependence of the dielectric constant of Al/NNO/SRO/(001)MgO heterostructure. \label{fig:Diel_All}}
\end{figure}

Our electric measurements indicate the presence of a ferroelectric phase at room temperature and a transition upon heating, as evidenced from ferroelectric hysteresis loops and the maximum in the temperature dependence of the dielectric constant. Figures~\ref{fig:Diel_All}(a)-(d) show dielectric constant dependencies of the Al/NNO/SRO/(001)MgO heterostructure on applied electric field $\varepsilon(E)$ measured at various frequencies $f_{\rm ext}$ of external triangular signal. At room temperature the NaNbO$_3$ film has a dielectric constant of $\varepsilon\sim900$ and a rather low loss tangent $\tan\delta\sim0.02 - 0.11$. Independent of $f_{\rm ext}$ the $\varepsilon(E)$ dependencies have a butterfly form characteristic of ferroelectric structures, possess hysteresis, and have high dielectric tunability $[\varepsilon(E=0~{\rm kV/cm})-\varepsilon(E=100~{\rm kV/cm})]/\varepsilon(E=0~{\rm kV/cm})\approx0.5$.

Figures~\ref{fig:Diel_All}(e) and (f) show electric polarization and electric current at different amplitudes of applied electric field. The dependencies possess characteristic ferroelectric form, which suggests that the NaNbO$_3$ film at room temperature in the studied heterostructure is in the ferroelectric state. Table~\ref{tab:FE_params} provides the main ferroelectric properties of the NNO film calculated from the $P(E)$ loop for the electric field amplitude of 340~kV/cm. The asymmetric form of the $P(E)$ loops reveals itself in the differences of coercive fields and remanent polarizations, which suggests the presence of internal field in the NNO film.


The temperature dependence of the dielectric constant shown in Fig.~\ref{fig:Diel_All}(g) reveals a broad maximum, whose position is independent on frequency. On heating, the dielectric maximum occurs at $T_m=500$~K, whereas on cooling at $T_m=440$~K. Thus, according to the temperature dependence of the dielectric constant, the NaNbO$_3$ film experiences a phase transition in the temperature range 440 -- 500~K characterized by large thermal hysteresis. In order to shed light on the origin of this phase transition we performed temperature-dependent x-ray reciprocal space mapping and Raman spectroscopy studies, the results of which are described in the following sections.

\begin{table}
\caption{\label{tab:FE_params}Ferroelectric $P(E)$ loop parameters for the 340~kV/cm electric field amplitude}
\begin{ruledtabular}
\begin{tabular}{lr}
Parameter & Value \\
\hline
$E_{c+}$, kV/cm & 111.6 \\
$E_{c-}$, kV/cm & -78.04 \\
$P_{r+}$, $\mu$C/cm$^2$ & 11.4 \\
$P_{r-}$, $\mu$C/cm$^2$ & -16.4 \\
$P_{rrel+}$, $\mu$C/cm$^2$ & 11.21 \\
$P_{rrel-}$, $\mu$C/cm$^2$ & -15.9 \\
$P_{max+}$, $\mu$C/cm$^2$ & 24.3 \\
$P_{max-}$, $\mu$C/cm$^2$ & -24.3 \\
$W_{loss}$, $\mu$J/cm$^2$ & 809 
\end{tabular}
\end{ruledtabular}
\end{table}

\subsection{Reciprocal space imaging}

\subsubsection{Room temperature}

\begin{figure}[t]
\centering
\includegraphics[width=\columnwidth]{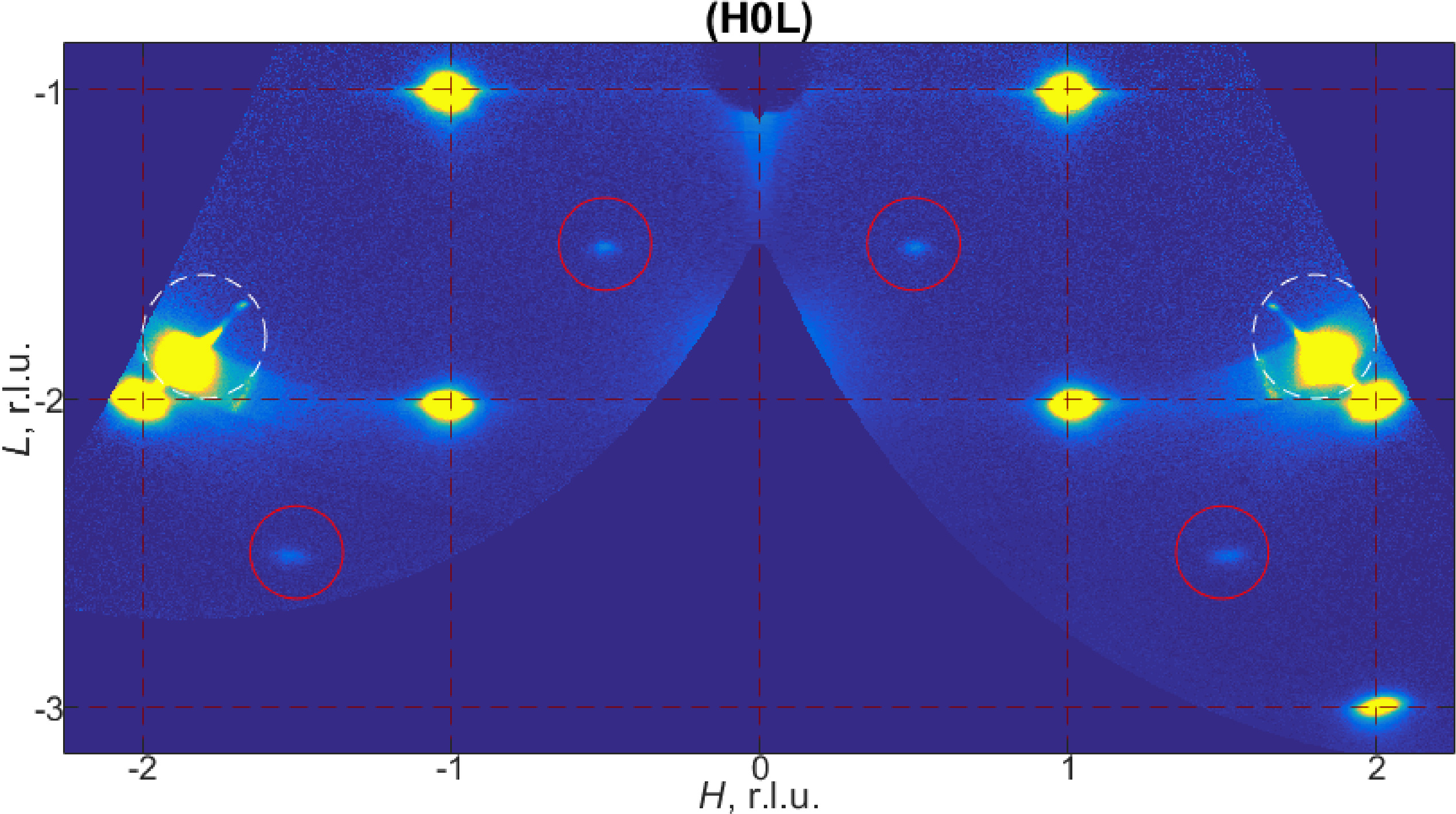}%
\caption{
(H0L) plane of reciprocal space for the NNO/SRO/MgO heterostructure at room temperature. Intensity peaks at $(H + \frac{1}{2}, 0, L + \frac{1}{2})$ are superstructural reflections from NNO (marked by red circles). Intensity peaks at $\approx(\pm1.8, 0, -1.8)$ are MgO reflections (marked by white circles).\label{fig:RSM1}}
\end{figure}

To study the crystal structure of sodium niobate we performed x-ray reciprocal space mapping. Three sets of reflections can be expected  in diffraction from the NNO/SRO/MgO heterostructure, i.e., one from the substrate and two from the NNO and SRO films, however only those from NNO and MgO are observed due to the small thickness of the intermediate layer SrRuO$_3$. The pseudocubic reciprocal space grid is based on the so-called main reflections from NNO, which is the object of the study (Fig.~\ref{fig:RSM1}).

\begin{figure}[t]
\centering
\includegraphics[width=\columnwidth]{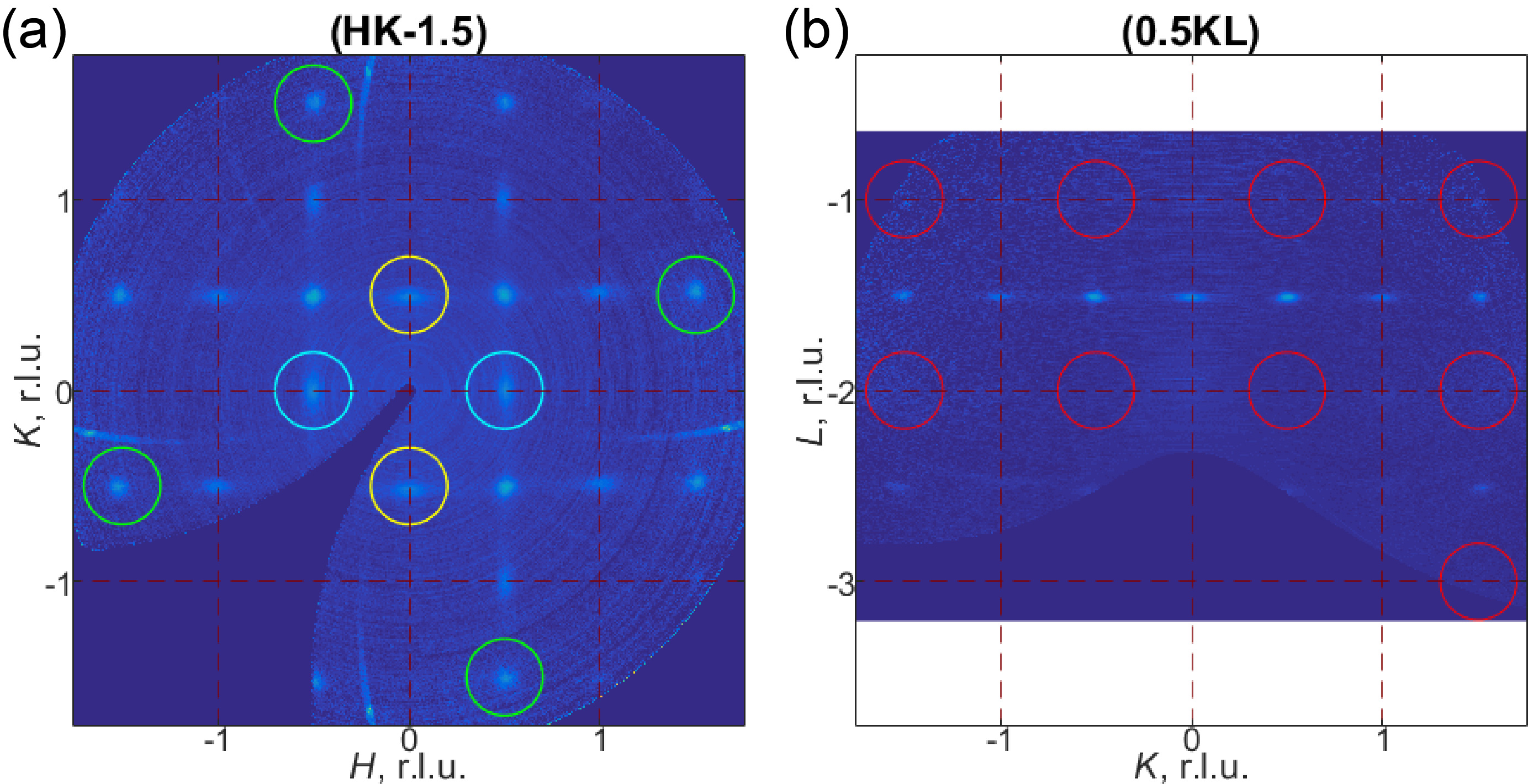}%
\caption{Room temperature reciprocal space maps of the NNO/SRO/MgO heterostructure. (a) $(HK-1.5)$ plane: superstructural reflections at M$_H$, M$_K$, and R positions of the PCBZ are marked by yellow, cyan, and green circles, respectively. (b) $(0.5KL)$ plane: almost vanishing superstructural reflections at M$_L$ positions marked by red circles can be seen.\label{fig:RSM2}}
\end{figure}

In addition to reflections in $\Gamma$ $(HKL)$ positions of the pseudocubic Brillouin zone (PCBZ), superstructural reflections in R- and M-points are found, which are shown in Fig.~\ref{fig:RSM2}(a). The family of M-points consists of reflections in M$_H=(H, K + \frac{1}{2}, L + \frac{1}{2})$, M$_K=(H + \frac{1}{2}, K, L + \frac{1}{2})$, and M$_L=(H + \frac{1}{2}, K + \frac{1}{2}, L)$ positions. At room temperature the reflections in M$_L$ points are very weak, as can be seen in Fig.~\ref{fig:RSM2}(b).

\begin{figure}[t]
\centering
\includegraphics[width=\columnwidth]{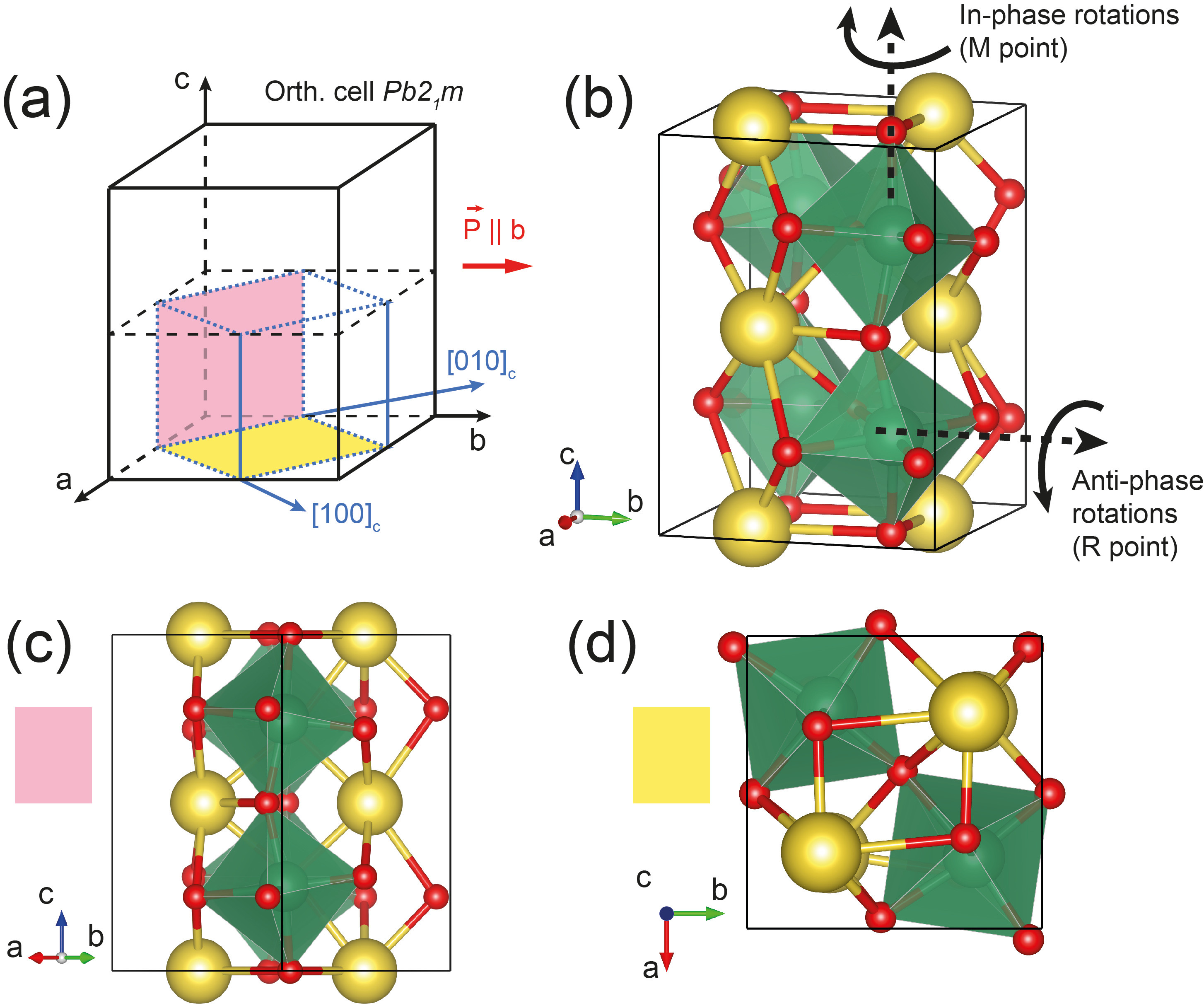}%
\caption{(a) The orthorhombic unit cell of the Q phase (black) and the constrained pseudocubic faces (shaded yellow and pink). The pseudocubic cell is shown in blue. Red arrow gives the direction of electric polarization. (b) Orthorhombic unit cell of the Q phase. The dark yellow and red spheres represent Na and O atoms, respectively, whereas NbO$_6$ octahedra are shown in green. (c) Top view of the "pink"-type epitaxy. (d) Top view of the "yellow"-type epitaxy. \label{fig:Pink_Yellow}}
\end{figure}

The set of observed superstructural reflections, taking into account their presence or absence in certain places of the reciprocal space, makes it possible to characterize the distorted structure. The first point to note is that M point reflections are observed only in points with symmetric coordinates, i.e., when the two half-integer coordinates are equal in magnitude. Such a condition for systematic absence of reflections exists only for two distortion modes in M point, namely, for the case of M$^{2+}$ (in-phase tilting of octahedra) and M$^{3+}$ (distortions of octahedra). Here and in the following we assume that Na is in the origin of the pseudocubic unit cell. Assuming that distortions are unlikely, because they are energetically less favorable, one can conclude that the M point reflections are due to in-phase tilting of octahedra. Similarly, since no reflections are observed at symmetric (all coordinates equal in magnitude) R points, which is only compatible with the R$^{5-}$ mode, the corresponding structural changes are due to anti-phase tilting of octahedra. Therefore, from reciprocal space mapping one can conclude that both in-phase and anti-phase tilting of octahedra are observed 
in the film at room temperature. This is compatible with the ferroelectric Q phase as shown in Figs.~\ref{fig:Pink_Yellow}(a) and (b).

\subsubsection{Temperature-dependent studies}

\begin{figure}[t]
\centering
\includegraphics[width=\columnwidth]{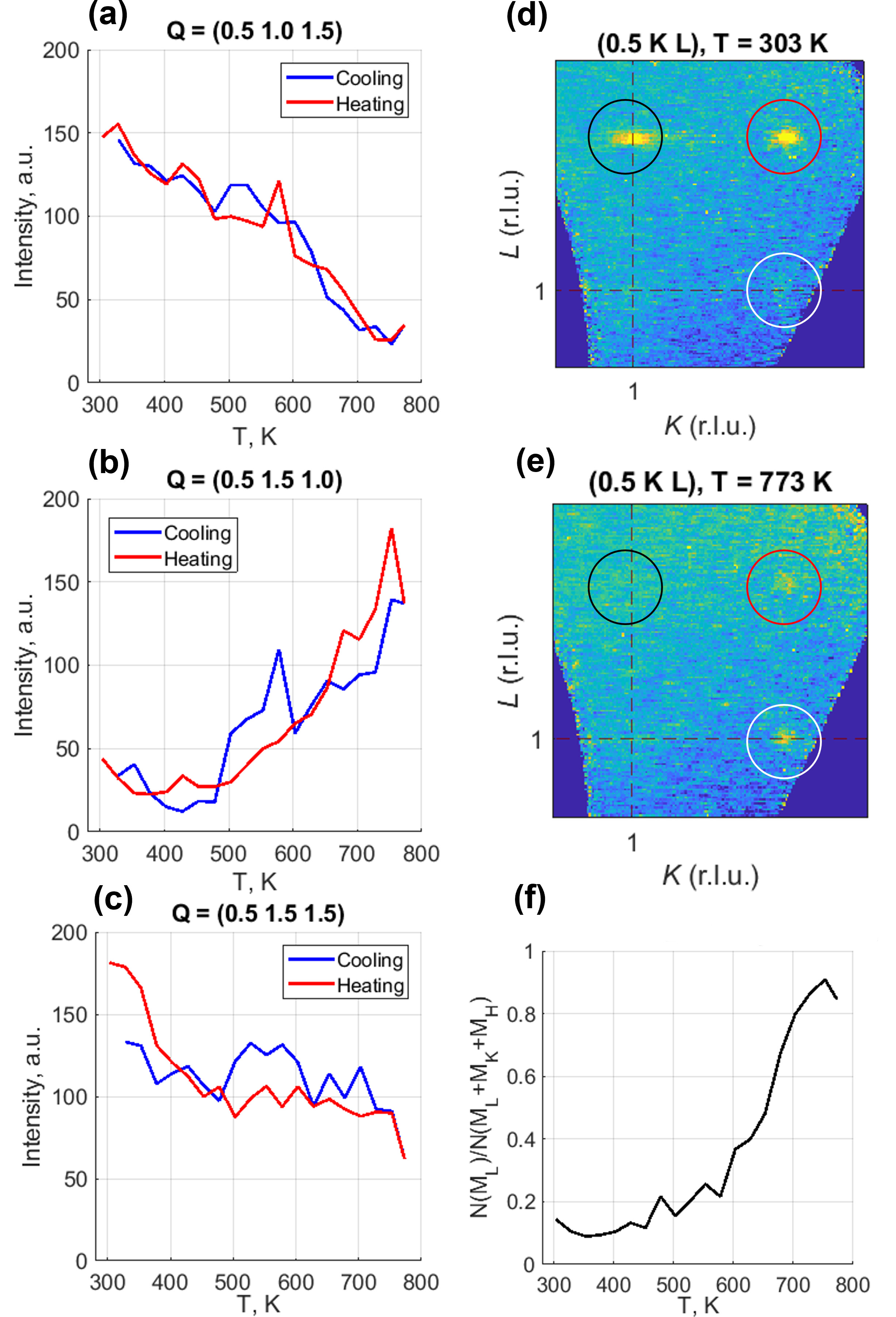}%
\caption{
Intensities of superstructural reflections at the M$_K$ $(\frac{1}{2}, 1, \frac{3}{2})$ point corresponding to in-phase rotations of octahedra around the axis parallel to the substrate (a), at the M$_L$ $(\frac{1}{2}, \frac{3}{2}, 1)$ point corresponding to the in-phase rotations of octahedra around the axis normal to the substrate (b), and at the R $(\frac{1}{2}, \frac{3}{2}, \frac{3}{2})$ point corresponding to anti-phase rotations (c). (d-e) $(\frac{1}{2}KL)$ planes at different T (M$_K$, M$_L$, and R points are marked by black, white, and red circles, respectively). (f) Volume fraction of domains with reflections in the M$_L$-points.\label{fig:RSM3_T_dep}}
\end{figure}

In order to study the evolution of the structure with temperature we performed reciprocal space mapping in the temperature range 298 -- 773~K. Upon heating, starting from room temperature a gradual extinction of superstructural reflections in M$_H$ and M$_K$ points occurs as shown in Fig.~\ref{fig:RSM3_T_dep}(a). These reflections do not disappear completely in the studied temperature range, as can be seen in Fig.~\ref{fig:RSM3_T_dep}(d,e). The same holds for the intensities in the R-point however to a lower extent, as shown in Fig.~\ref{fig:RSM3_T_dep}(c). On the other hand, with heating the intensities at the M$_L$-points increase as shown in Fig.~\ref{fig:RSM3_T_dep}(b). The results of the temperature-dependent studies indicate occurrence of a diffuse phase transition between two different structures, which starts at room temperature and extends at least to the highest measured temperature (773~K). This phase transition can be interpreted as between two different orientations of the Q phase, which we denote by "pink" and "yellow" shaded faces in Fig.~\ref{fig:Pink_Yellow}(a) that have to be constrained to the underlying substrate as shown in Figs.~\ref{fig:Pink_Yellow}(c) and (d). The details of theses structures will be discussed below. Based on the intensities of reflections in M$_H$, M$_K$, and M$_L$ points one can calculate the temperature evolution of the relative ratio of different phases, which is shown in Fig.~\ref{fig:RSM3_T_dep}(f).

\subsection{Raman spectroscopy}

\begin{figure}[t]
\centering
\includegraphics[width=\columnwidth]{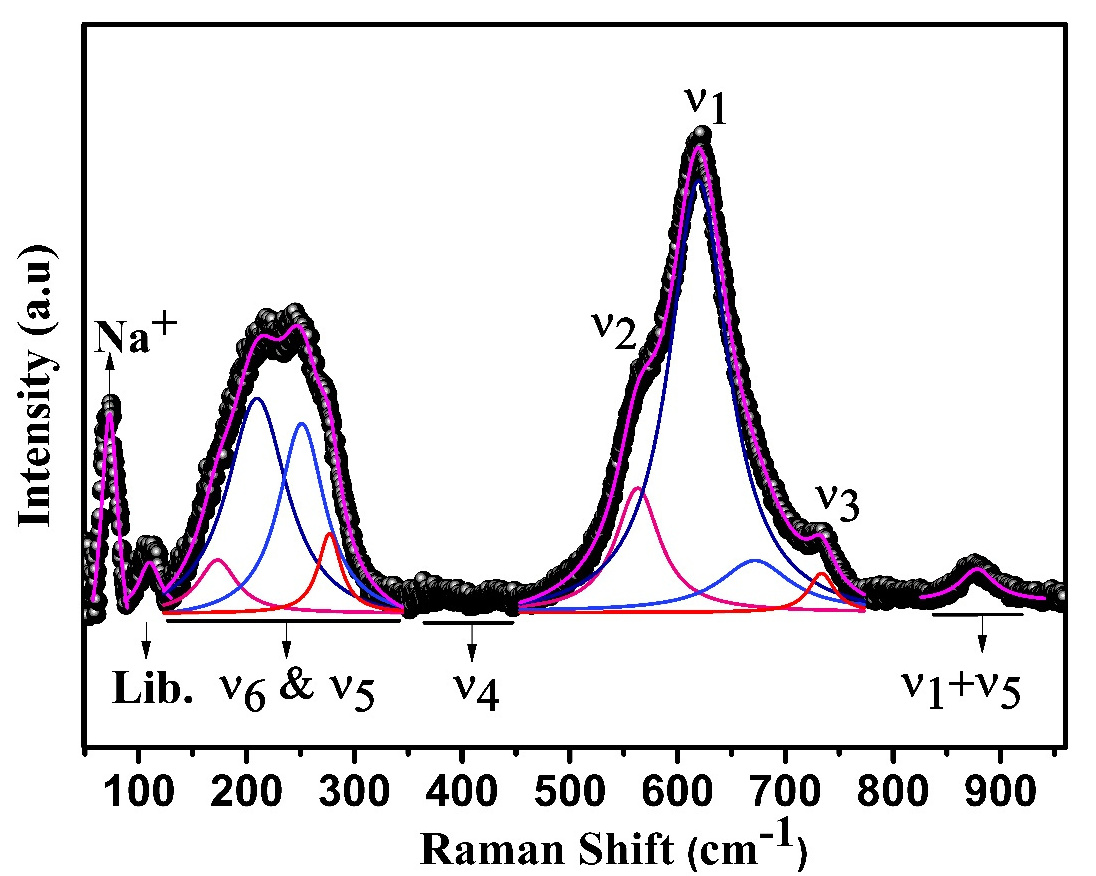}%
\caption{
Raman spectrum of NNO/SRO/MgO thin film at room temperature with mode assignment showing the translational mode of Na$^+$, the librational mode, and the modes $\nu_1$ -- $\nu_6$ that stem from vibrations of a NbO$_6$ octahedron with O$_h$ symmetry.\label{fig:Raman_1}}
\end{figure}

To further understand how the structure changes with the temperature, we have carried out temperature-dependent Raman Spectroscopy on the NNO/SRO/MgO heterostructure from 298 to 853~K. The Raman spectrum recorded at ambient conditions is shown in Fig.~\ref{fig:Raman_1} and the modes are assigned as the translational modes of Na$^+$ cation, librational modes, and internal vibrations of NbO$_6$ octahedra ($\nu_1$ -- $\nu_6$) based on the earlier reported studies on NaNbO$_3$ powder, single crystals, and ceramics~\cite{Raman1,Raman2,Raman3,Raman4,Raman5,Raman6,Shiratori_SizeEffect_2005,Yuzyuk_INC_NaNbO3_2005,Raman9}. Here the modes $\nu_1$ -- $\nu_6$ stem from those of an ideal undistorted NbO$_6$ octahedron with O$_h$ symmetry~\cite{Ross_1970}. The spectra at room temperature mainly consist of two broad asymmetric bands, and both were fitted into four Lorentz peaks each. As shown in Fig.~\ref{fig:Raman_1}, the $\nu_1$, $\nu_2$, and $\nu_3$ phonon modes of NbO$_6$ octahedral vibrations were seen together as the most intense band around 450-780~cm$^{-1}$ and the $\nu_5$ and $\nu_6$ bands were found to be converged as a broad, second most intense band ranging from 130 to 340~cm$^{-1}$. The $\nu_4$ modes were observed to be very weak and the weak band at 877~cm$^{-1}$ is assigned to the $\nu_1+\nu_5$ combination mode. In addition, there are two relatively sharp peaks observed around 73~cm$^{-1}$ and 110~cm$^{-1}$, which are assigned to the external vibration related to the translational movements of Na$^+$ cations and the librations of the NbO$_6$ octahedra, respectively~\cite{Raman1,Raman2,Raman3,Raman4,Raman5,Raman6,Shiratori_SizeEffect_2005,Yuzyuk_INC_NaNbO3_2005,Raman9}. As compared to the Raman spectra of bulk NaNbO$_3$ in P phase, remarkable differences were observed in our thin film spectra such as the reduced number of modes (especially the external vibrations of Na$^+$ cations and librational modes of NbO$_6$ octahedra) and the increased broadness of the internal octahedral vibrations ~\cite{Raman1,Raman2,Raman3,Raman4,Raman5,Raman6,Shiratori_SizeEffect_2005,Yuzyuk_INC_NaNbO3_2005,Raman9,Yuzyuk_MgO_film_2010}. Similar observations in Raman spectra were also seen in the Q phase reported in NaNbO$_3$ powders with particle sizes below 400~nm~\cite{Shiratori_SizeEffect_2005}. As compared to the NNO single crystal, the absence of modes around 96 and 154~cm$^{-1}$, the overlapping modes in the broad bands and the shifting of modes from 557 and 601~cm$^{-1}$ to 564 and 620~cm$^{-1}$, respectively, suggest that the phase occurring in the NNO thin film is the ferroelectric Q phase, similar to that of the earlier reported ceramic and thin film samples~\cite{Yuzyuk_INC_NaNbO3_2005,Yuzyuk_MgO_film_2010}. This conclusion is correlating well with our results from X-ray reciprocal space imaging too.

\begin{figure}[t]
\centering
\includegraphics[width=\columnwidth]{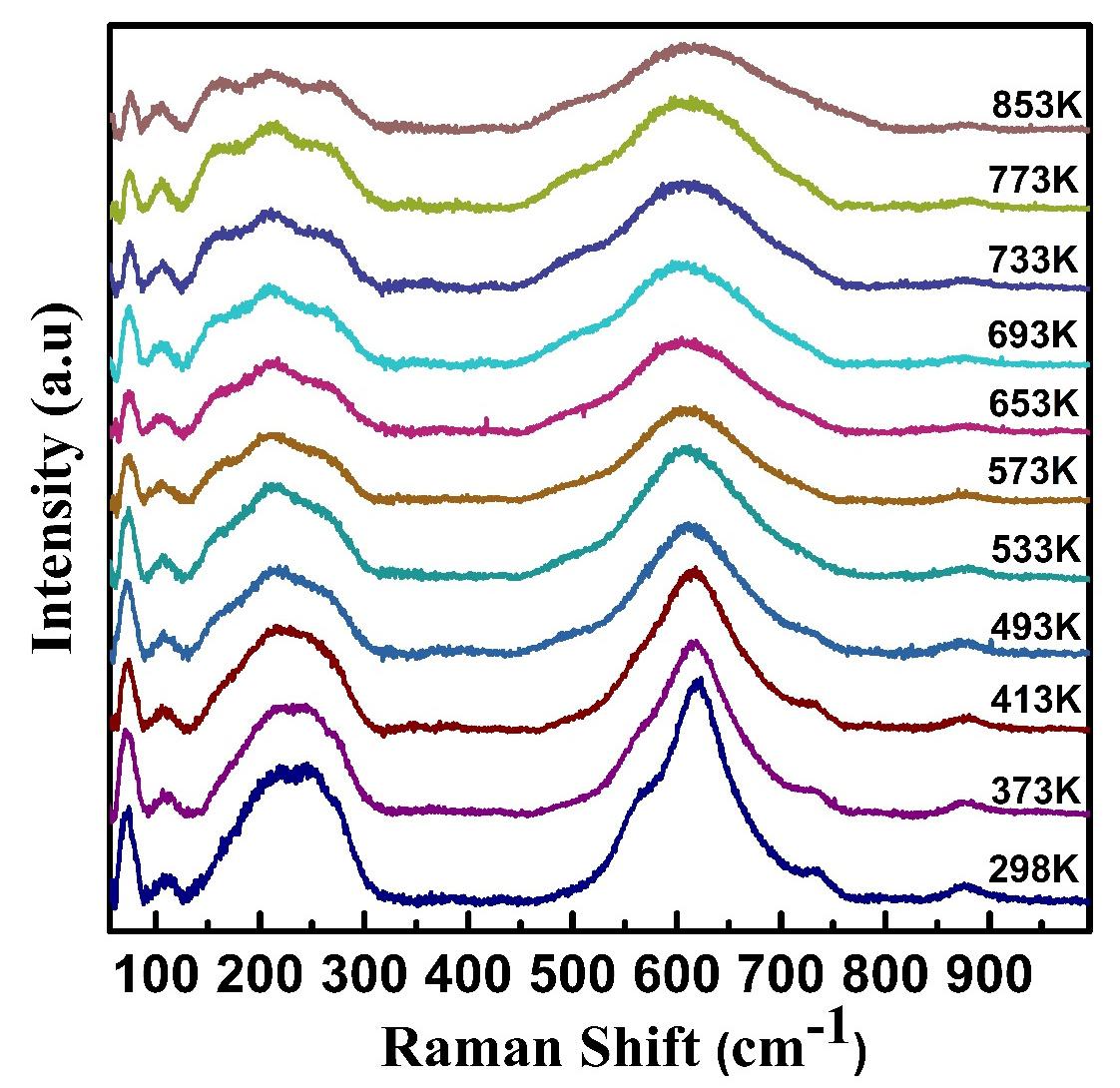}%
\caption{
The temperature-dependent evolution of Raman spectra at specific temperature steps.\label{fig:Raman_2}}
\end{figure}

The temperature evolution of Raman spectra of NNO thin film at specific temperature steps are shown in Fig.~\ref{fig:Raman_2}. It can be noticed that in general, there is a broadening of peaks with increase in temperature as expected and a change in the relative intensity of the peaks. The distinctive Raman spectra of the Q phase similar to the ambient Raman spectrum were seen in the temperature range of 298 -- 400~K. However, with further increase in temperature, there are gradual changes observed in the spectra between 400 and 500~K. While four Raman modes could be seen ranging from 130 to 340~cm$^{-1}$ until 400~K, we were able to fit only three peaks above 400~K. From this temperature the fitting procedure indicates, that the two modes around 248 and 275~cm$^{-1}$ merge together and a new mode is seen at 260~cm$^{-1}$, which is accompanied by an abrupt change in the position and intensity of the other modes in the phonon bands ($\nu_5$ and $\nu_6$). This new mode continues to exist till higher temperatures. Also, the $\nu_3$ mode at 671~cm$^{-1}$ gradually decreases in intensity and disappears into the background after 400~K.

\begin{figure}[t]
\centering
\includegraphics[width=\columnwidth]{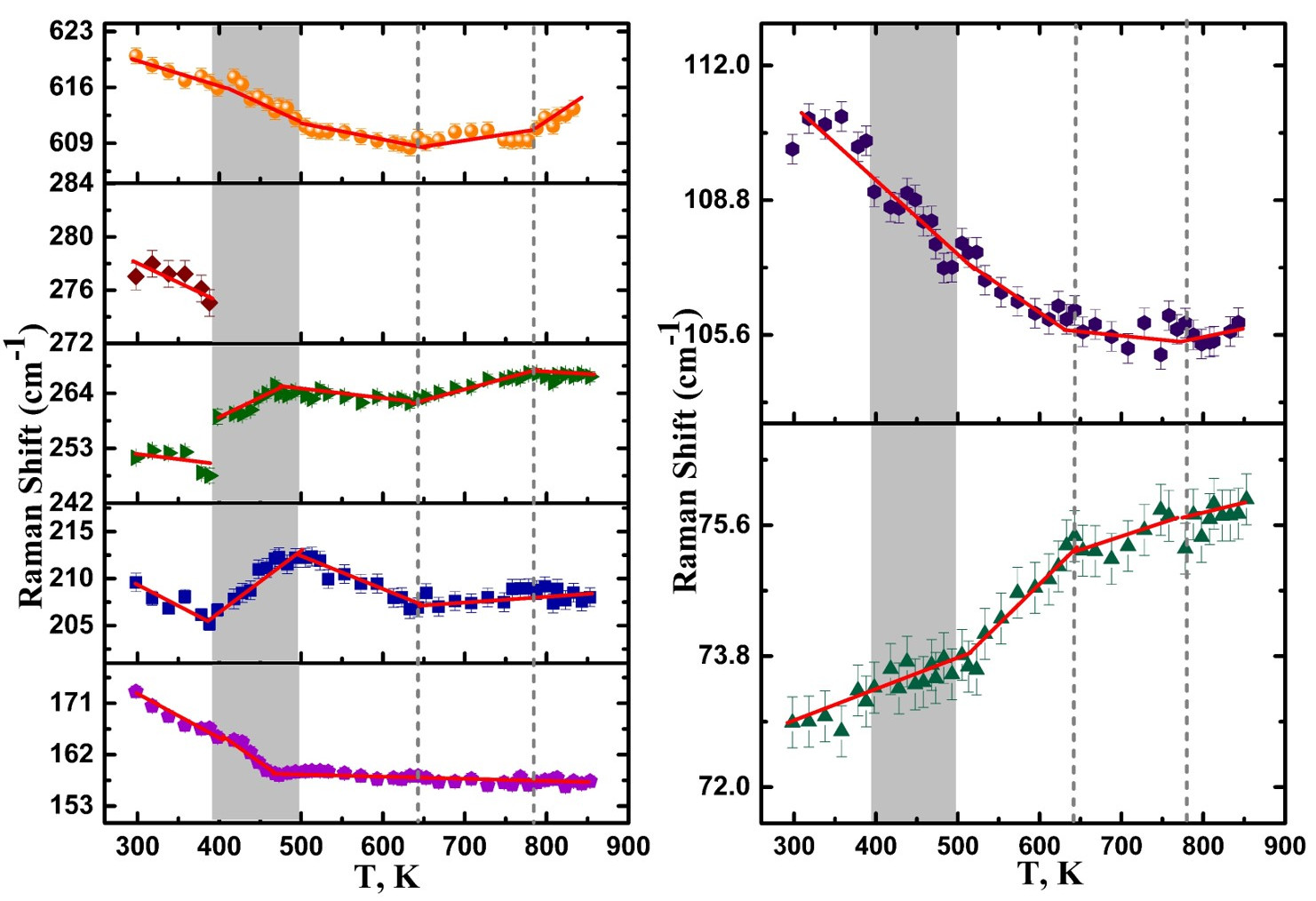}%
\caption{
Raman peak shift vs. temperature plots for different Raman modes during heating. The grey coloured region indicates the region of phase transition, while the dashed lines show possible additional anomalies. The solid red lines are guide to the eye.\label{fig:Raman_3}}
\end{figure}

In order to clearly understand the variation in the spectra with the change of temperature, the peak positions of some of the intense Raman modes have been plotted with the increasing temperature, as shown in Fig.~\ref{fig:Raman_3}. Up to 400~K, all the phonon modes of NbO$_6$ octahedra show softening with increase in temperature, as expected. From 400~K, because of the rearrangement of modes, the two peaks around 210 and 260~cm$^{-1}$ start to harden while the modes at 170 and 620~cm$^{-1}$ [Nb-O symmetric stretching ($\nu_1$) mode] continue to soften with small slope changes~\cite{Raman4}. On further heating, all the internal vibrational modes of the octahedra show another anomaly and start softening above 500~K except the 170~cm$^{-1}$ mode which does not undergo any significant changes. It is important to note that, as opposed to the expected behaviour, the 73~cm$^{-1}$ peak assigned as the translational mode of Na$^+$ cation shows hardening in the entire temperature range while the librational mode at 110~cm$^{-1}$ softens. We could also observe that most of the Raman modes show minor slope changes around 650 and 780~K.

\section{Discussion}


The $P(E)$ hysteresis loop measurements as well as the form of the $\varepsilon(E)$ curves reveal that sodium niobate in the NNO/SRO/(001)MgO heterostructure is in polar and switchable state and the electric polarization component exists along the normal to the substrate. However, the true direction of polarization can not be unambiguously determined from electric measurements because of the parallel geometry of electrodes, which can not exclude possible in-plane component of polarization. 

It is reasonable to look for the candidates for the phases of NaNbO$_3$ observed at room and elevated temperatures in this work among those found in the bulk form of NNO or the phases related to them. The influence of epitaxial strain in (001)NaNbO$_3$ films has been studied using the DFT method in the recent work by Patel {\it et al.}~\cite{Pattel_2021}. Their results indicate that at compressive strains and tensile strains up to $\approx1.27$\% the monoclinically distorted N phase having the $Cc$ symmetry is stabilized. In the narrow range of tensile strains between $\approx1.27$\% and $\approx1.5$\% the phase with $Pca2_1$ symmetry is found, which originates from the P bulk phase ($Pbcm$) by additional emergence of electric polarization along the $[001]$ orthorhombic direction. At tensile strains above $\approx1.5$\% the Q phase is found to possess the lowest energy.

Out of these three possible polar structures, namely the $Cc$, $Pca2_1$, and Q phases, the first two can be excluded from considerations because they contradict our x-ray observations. The $Cc$ phase, similar to the bulk N phase, is characterized by distortions in the R-point of the PCBZ, but contrary to our results does not experience M-point distortions~\cite{Toledano_AF_2019}. The $Pca2_1$ phase should show reflections in the $\Delta$ $(0,0,1/4)$ point of the PCBZ, which we do not observe in our experimental data. Therefore, among the obvious candidates only the Q phase remains potentially possible.

The Q phase appears due to the combination of the R, M, and $\Gamma$-point instabilities~\cite{Cochran_Zia_1968}, which corresponds to our observations in reciprocal space imaging. Figure~\ref{fig:Pink_Yellow}(a) shows schematical view of the unit cell of the Q phase and its relation to the pseudocubic unit cell, whereas Fig.~\ref{fig:Pink_Yellow}(b) shows the corresponding view of the atomic positions and NbO$_6$ octahedra with indicated rotations due to the modes in the R and M points. Each domain of the bulk Q phase is characterized by a single M-point reflection, i.e., by either of M$_H$, M$_K$, or M$_L$ point. In Fig.~\ref{fig:Pink_Yellow}(a) the M$_L$ point is active if we assume that the normal to the substrate is parallel to the vertical direction in the figure. At room temperature we observe reflections in the M$_H$ and M$_K$ points, which should stem from differently oriented domains of the studied structure. Considering the orthorhombic unit cell of the Q phase, one can note that there can be two different epitaxial contacts between the NaNbO$_3$ film and the (001)SrRuO$_3$/MgO substrate. These two different epitaxial contacts are denoted by pink and yellow shaded faces as shown in Fig.~\ref{fig:Pink_Yellow}(a), that should be constrained to the underlying SrRuO$_3$ layer. The respective top views of the structures are shown in Figs.~\ref{fig:Pink_Yellow}(c) and (d). One should note here, that in the theoretical work of Patel {\it et al.}~\cite{Pattel_2021} only the "yellow"-type epitaxial contact has been considered, whereas the "pink" one has been overlooked.

Having identified the Q phase with "pink" epitaxial contact as the room temperature structure, we now come to the question on how this structure changes upon heating and what is the origin of the broad peak in the dielectric constant. Upon increasing temperature, the diffraction pattern gradually changes due to appearance of reflections in the M$_L$-type points and extinction of the M$_H$ and M$_K$ points. The structure with M$_L$-type reflections corresponds then to the "yellow"-type epitaxial contact with electric polarization parallel to the substrate, as shown in Fig.~\ref{fig:Pink_Yellow}(d). However, the presence or absence of electric polarization in the high-temperature phase can not be judged upon based on the available experimental data, because our x-ray diffraction results are consistent with the "yellow" Q phase epitaxy, but can not confirm the presence of polar distortions. In principle the high-temperature structure can be characterized only by R and M point distortions, which may correspond to the bulk T$_1$ phase~\cite{Toledano_AF_2019}. However, the possibility of the T$_1$ phase option seems to us less likely since this phase appears in the bulk at substantially higher temperatures.

The variant with temperature-driven change of Q phase epitaxy between the "pink" and "yellow" orientations is interesting because it presents a rare type of phase transitions in films, which was extensively discussed in theoretical works earlier, and at the same time can explain the observed broad maximum of the dielectric constant. Such a phase transition between two orthorhombic phases with differently oriented polarizations has not been studied according to our knowledge. However, a phase transition in PbTiO$_3$ films between two tetragonal phases $a$ and $c$ with in-plane and out-of-plane orientations of polarization, respectively, has been studied theoretically by Pertsev {\it et al.}~\cite{Pertsev_Films_2001}. The results indicate that the dielectric constant experiences a maximum across this phase transition. Thus, one may expect similar behavior across the phase transition between two orthorhombic phases observed in our case.

The results from the dielectric measurements and the reciprocal space mapping are further corroborated by the Raman spectroscopy data, which suggest the presence of Q phase structure at room temperature. The Raman spectroscopy data reveal changes in the Raman spectra at 400 and 500~K, which, thus, supports the results of the other measurement methods showing a broad and diffuse phase transition in this temperature range. The anomalies in the Raman spectra at 400 and 500~K consist in the disappearance of certain modes upon heating and slope changes in the peak positions of modes when plotted against temperature. This disappearance of modes is consistent with the change of Q phase orientation from "pink" to "yellow". As our sample is a highly oriented thin film, any change in the orientation of the orthorhombic Q phase cell should be reflected in the Raman spectra taken in the same backscattering geometry. The reduced number of modes observed above 400~K is in qualitative correspondence with the change in epitaxial contacts of the Q phase because of the Raman scattering selection rules for the two considered orientations. Indeed, in the case of the "pink"-type epitaxial contact, the employed backscattering geometry of the experiment allows the modes A$_1$(LO), A$_1$(TO), A$_2$, and B$_1$(TO), whereas the "yellow"-type contact allows only the A$_1$(TO) and B$_2$(TO) modes~\cite{aroyo2006bilbao1,aroyo2006bilbao2}.

In contrast to x-ray and dielectric measurements, the changes observed in the Raman spectroscopy are more abrupt. However, one has to keep in mind the different scales at which the information is gathered in these methods. In the case of Raman measurements the dimensions of the laser spot is of the order of several microns and the spectra are averaged over a low number of different domains, whereas in the case of x-ray and dielectric methods the dimensions are of the order of hundreds of microns, which results in averaging over many different domains. Furthermore, although we also observe some slope changes in Raman spectra on further heating around 650 and 780~K, the exact nature of these anomalies could not be proposed from the Raman data alone. They might still reflect the gradual changes observed in reciprocal space mapping up to the highest measured temperature (773~K), since the anomalies in the Raman spectra fall into the temperature range covered by the x-ray diffraction.

\section{Conclusions}

Using the combination of the dielectric, x-ray diffraction, and Raman spectroscopy measurement methods we showed that NaNbO$_3$ film in  the synthesized NaNbO$_3$/SrRuO$_3$/(001)MgO heterostructure experiences a diffuse phase transition above room temperature. Our results allow concluding that this transition is related to the transition between two different orientations of the Q phase, that have different epitaxial conjugations with the underlying SrRuO$_3$/(001)MgO structure. This phenomenon is remarkably interesting because such phase transitions have attracted attention earlier in theoretical works~\cite{Pertsev_Films_1998,Pertsev_films_1999_equil,Pertsev_Films_2001,Pertsev_PZT_2003}, however their experimental characterization is limited. The discovered behavior of NNO films can be potentially interesting for the development of applications in memory, logic, sensing, and energy harvesting devices.

\begin{acknowledgments}
C.N., A.J., and J.S performed Raman studies and acknowledge JNCASR/TRC for the Raman Spectrometer facility. A.J. and J.S. acknowledge Jawaharlal Nehru Centre for Advanced Scientific Research for providing research fellowships (JNC/S0752 and JNC/S0539), which supported Raman measurements and data treatment. M.V.V., A.E.G., S.A.U., and R.G.B. performed x-ray diffraction studies. A.V.P., D.V.S., I.P.R., and N.V.T. have conceptualized the study, performed synthesis, dielectric measurements, and collaborative interpretation of results under the financial support by the Russian Science Foundation grant No. 19-12-00205.
\end{acknowledgments}



%

\end{document}